\documentclass[conference]{IEEEtran}
\IEEEoverridecommandlockouts
\usepackage{cite}
\usepackage{array}
\usepackage{amsmath,amssymb,amsfonts}
\usepackage{algorithmic}
\usepackage{graphicx}
\usepackage{textcomp}
\usepackage{xcolor}
\usepackage{booktabs}
\usepackage{multirow}
\usepackage{tabularx}
\usepackage{etoolbox}
\makeatletter
\patchcmd{\@makecaption}
  {\scshape}
  {}
  {}
  {}
\makeatletter
\patchcmd{\@makecaption}
  {\\}
  {.\ }
  {}
  {}
\makeatother
\def\tablename{Table}

\def\BibTeX{{\rm B\kern-.05em{\sc i\kern-.025em b}\kern-.08em
    T\kern-.1667em\lower.7ex\hbox{E}\kern-.125emX}}
\begin{document}

\title{Graph Neural Networks Enhanced Smart Contract Vulnerability Detection of Educational Blockchain\\
\thanks{Corresponding author: Zhifeng Wang, Email: zfwang@ccnu.edu.cn}
}

\author{\IEEEauthorblockN{1\textsuperscript{st} Zhifeng Wang}
\IEEEauthorblockA{\textit{CCNU Wollongong Joint Institute} \\
\textit{Central China Normal University}\\
Wuhan 430079, China \\
zfwang@ccnu.edu.cn}
\and
\IEEEauthorblockN{2\textsuperscript{nd} Wanxuan Wu}
\IEEEauthorblockA{\textit{CCNU Wollongong Joint Institute} \\
	\textit{Central China Normal University}\\
	Wuhan 430079, China \\
	wuwx0428@163.com}
\and
\IEEEauthorblockN{3\textsuperscript{rd} Chunyan Zeng}
\IEEEauthorblockA{\textit{School of Electrical and Electronic Engineering} \\
	\textit{Hubei University of Technology}\\
	Wuhan 430068, China \\
	cyzeng@hbut.edu.cn}
\and
\IEEEauthorblockN{4\textsuperscript{th} Jialong Yao}
\IEEEauthorblockA{\textit{CCNU Wollongong Joint Institute} \\
	\textit{Central China Normal University}\\
	Wuhan 430079, China \\
	jy123@uowmail.edu.au}
\and
\IEEEauthorblockN{5\textsuperscript{th} Yang Yang}
\IEEEauthorblockA{\textit{CCNU Wollongong Joint Institute} \\
	\textit{Central China Normal University}\\
	Wuhan 430079, China \\
	univeryang@ccnu.edu.cn}
\and
\IEEEauthorblockN{6\textsuperscript{th} Hongmin Xu}
\IEEEauthorblockA{\textit{Faculty of Artificial Intelligence in Education} \\
	\textit{Central China Normal University}\\
	Wuhan 430079, China \\
xhm@ccnu.edu.cn}
}

\maketitle

\begin{abstract}
With the development of blockchain technology, more and more attention has been paid to the intersection of blockchain and education, and various educational evaluation systems and E-learning systems are developed based on blockchain technology. Among them, Ethereum smart contract is favored by developers for its ``event-triggered" mechanism for building education intelligent trading systems and intelligent learning platforms. However, due to the immutability of blockchain, published smart contracts cannot be modified, so problematic contracts cannot be fixed by modifying the code in the educational blockchain. In recent years, security incidents due to smart contract vulnerabilities have caused huge property losses, so the detection of smart contract vulnerabilities in educational blockchain has become a great challenge. To solve this problem, this paper proposes a graph neural network (GNN) based vulnerability detection for smart contracts in educational blockchains. Firstly, the bytecodes are decompiled to get the opcode. Secondly, the basic blocks are divided, and the edges between the basic blocks according to the opcode execution logic are added. Then, the control flow graphs (CFG) are built. Finally, we designed a GNN-based model for vulnerability detection. The experimental results show that the proposed method is effective for the vulnerability detection of smart contracts. Compared with the traditional approaches, it can get good results with fewer layers of the GCN model, which shows that the contract bytecode and GCN model are efficient in vulnerability detection.
\end{abstract}


\section{Introduction}
The education blockchain refers to the use of blockchain as technical support when carrying out reforms to the traditional education systems \cite{9468408,Lyu2022}. The white paper on blockchain technology released by China in 2016 states that ``the transparency and immutability of the blockchain system are perfectly suitable for student credit management, further education and employment, academics, qualification certification, and industry-academia cooperation, and are of great value to the healthy development of education and employment" \cite{9459573}. According to the visual analysis of the blockchain in education \cite{8247166,visualanalysis,Zeng2022}, blockchain technology is using its decentralized feature to break the absolute management power of traditional education administrators over education and promote the development of education in the direction of more equity \cite{9893128,Wang2022ac}. With the creation and development of Ethernet smart contract technology, programs can be implemented to automatically execute without third-party intervention after meeting the conditions to achieve functions such as controlling the assets of the blockchain and storing data information. By embedding smart contracts, blockchain technology can build virtual economy education intelligent transaction systems \cite{__2017}, which can promote the construction of a new system combining the Internet and education, avoid the limitations of the traditional education model in space and time to a certain extent, and help promote the change of the education system and accelerate its development.

Blockchain-based smart contract systems have many advantages, such as ensuring the authenticity \cite{Wang2022t} and security of information \cite{8246573,Zeng2022a}, saving human resources, improving the efficiency of program execution, etc. However, smart contracts are not absolutely secure. Different security vulnerabilities may exist throughout the life cycle of a smart contract, and due to the published code cannot be modified, the security problems caused by smart contract vulnerabilities will increase, so it is especially important to improve its security.

For example, the main detection in this paper is a timestamp dependency vulnerability. Smart contracts use timestamps to control certain important block control flow decisions, and if an attacker masquerades as a miner, they can bypass certain operations in the contract that are restricted by timestamps by maliciously controlling the range of timestamp generation.

With the continuous development of deep learning techniques, some scholars have proposed the use of these techniques for vulnerability detection to make it more accurate, comprehensive, and efficient. This paper uses Control Flow Graph (CFG) built based on bytecode files of smart contracts, use it as the input of a graph neural network, and builds a Graph Convolutional Network (GCN) model to realize vulnerability detection of smart contracts. The contributions of this paper can be summarized as follows:
\begin{itemize}
    \item A GCN model is built and successfully predicts contract vulnerabilities for the educational blockchain. 
    \item The vulnerabilities can be effectively detected through the bytecode files of smart contracts.
    \item The accuracy of model prediction can be increased if semantic processing is added or classification of edges is added.
\end{itemize}

The rest of the paper is organized as follows. In Section 2, we review the related work. In Section 3, we introduce the main research methods, including CFG composition and GCN model. In Section 4, we describe the details of the experiment and the results. Finally, we have a summary of this work in Section 5.

\section{Related Work}

We first introduce the contractual vulnerability detection methods that are now available. Then we summarize the development of graph convolutional neural networks.

\subsection{Contract Vulnerability Detection Methods}

In response to the security problems caused by smart contracts, numerous research teams at home and abroad have proposed solutions that seek to protect users' property security and data security. The current detection techniques mainly include two types, one is based on non-deep learning methods and the other is based on deep learning methods.

A non-deep learning-based method, the automated contract vulnerability mining tool Oyente \cite{10.1145/2976749.2978309}, is a symbolic execution-based analysis method. Using the bytecode file of a smart contract as input, after analyzing the bytecode and constructing the CFG, the Z3 solver is used to analyze the conditional jumps in the contract, which can predict whether there are seven types of vulnerabilities such as integer overflow errors and reentrant vulnerabilities for that contract.

Another non-deep learning-based approach, ContractFuzzer, is a fuzzy test-based detection tool. It consists of two parts, an offline EVM staking tool and an online fuzzy testing tool \cite{10.1145/3238147.3238177}. The tool generates legally valid inputs and mutated inputs that cross the valid boundary by analyzing the bytecode of the smart contract as well as the ABI interface; after starting the fuzzy test, the detection results of the contract can be obtained through the execution log.

A deep learning-based approach uses RNN networks. an RNN is a recurrent neural network that uses sequence data as input to a neural network, recursively in the direction of sequence data and with all recurrent units connected in a chain-like manner \cite{Zeng2021a}. In the RNN network model proposed in the literature \cite{__2021-1273}, two layers of threshold recursive units (GRUs) are connected after the embedding layer, and the fully connected layer is connected afterward. This experiment demonstrates that vulnerability detection can be done using smart contract operation sequences combined with deep learning networks.

Another deep learning-based method can use the Long Short-Term Memory Network (LSTM) \cite{Zeng2020}, a model that constructs three gates: input, output, and forgetting gates, implementing an optimization of the RNN and therefore providing further performance improvements. The use of this network model for vulnerability detection is proposed in the literature \cite{tann_towards_2019}, using a binary vector encoding representing the opcode of a smart contract as an input to the network model, and its experimental results show more effective detection results in contrast to non-deep learning methods.

\subsection{Graph Convolutional Network Model}

The graph neural network model used in this paper is the GCN. It is a model evolved from Spectral CNN and Chebyshev Network (ChebNet) \cite{TextClassification}. The important architecture of GCN includes a graph convolution layer, a graph readout layer, and a graph regularization layer to improve model generalization performance and a graph pooling layer to reduce the number of computational parameters. The GCN model is essentially the same as a Convolutional Neural Network (CNN) \cite{Zeng2022b}, i.e., it aggregates pro-domain information for operation, but the difference is that the GCN model applies to data with a non-Euclidean structure.

Because the GCN network deals with graph structures, it needs to be represented as multiple files during data preprocessing, such as adjacency matrix, number of nodes and information, number of edges and information, etc.

Since its introduction, the GCN network model has received a lot of attention from scholars from all walks of life and has been actively applied to various application sites of graph data. Currently, the GCN network model has been applied to the blockchain, biochemistry, traffic prediction, computer vision, and other fields with promising results \cite{Wang2021}.

\section{Proposed Method}
Smart contracts are run by EVM, which first compiles the source code into bytecode and then runs it as bytecode, so it is more realistic to use bytecode files as the basis for vulnerability detection.

Inspired by the success of deep learning in the fields of data mining \cite{Wang2023,Lyu2022,Li2023a,Min2019,Wang2022as}, computer vision \cite{Wang2021,Zeng2021c,Wang2022ac,Zeng2020,Li2023,Zeng2022,Wang2017,Zeng2022b,Wang2015a,Zeng2020a,Tian2018a,Min2018,Wang2022at} and speech processing \cite{Wang2021m,Zeng2021a,Wang2022t,Zeng2022a,Wang2020h,Zeng2021b,Wang2018a,Zeng2018,Wang2015b,Zhu2013,Wang2011,Wang2011a}, this paper proposes a method based on deep learning for smart contract vulnerability detection. We decompile the bytecode file of a smart contract to get the opcode, divide several basic blocks according to the instruction semantics and construct a control flow graph CFG according to the jump order. Then we build a GCN model including an input layer, several hidden layers, and an output layer structure, and then input data and test it. The overall experimental framework is shown in Fig. 1.

\begin{figure*}[!ht]
\label{fig:10Experiment structure}
\includegraphics[scale=0.265]{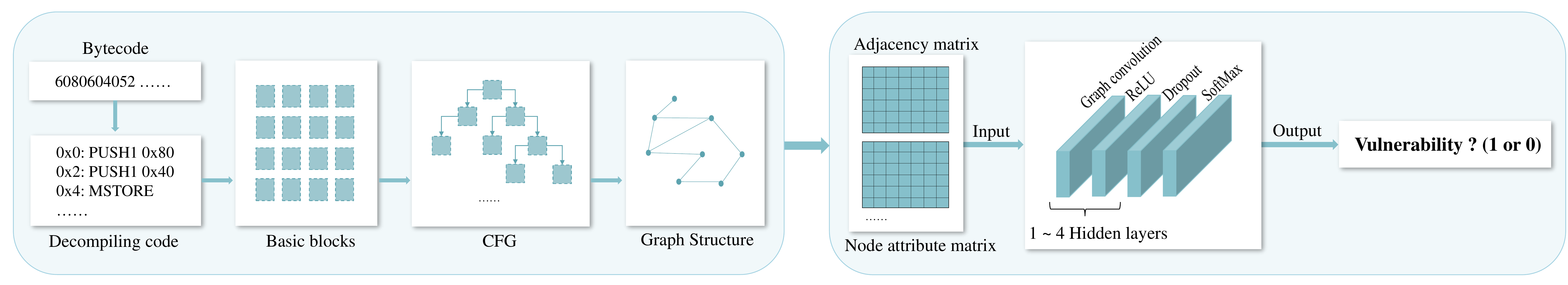}
\caption{A description of the experimental procedure. The first part is the process of constructing CFG, and the second part is the process of constructing GCN model and inputting data for vulnerability prediction.}
\end{figure*}
The process can be divided into the following main steps.
\begin{enumerate}
    \item Analyze smart contract bytecode files and generate decompiled code;
    \item Divide the basic blocks, add dependent edges to the basic blocks to build CFG, and use them as input to the GCN;
    \item Define the convolutional network layer of GCN;
    \item Add pooling layer, fully connected layer, etc. to build a complete GCN model.
\end{enumerate}

\subsection{Byte Code Analysis}

\subsubsection{Contract Bytecode Structure}
\begin{figure}[!ht]
\centering
\includegraphics[scale=0.49]{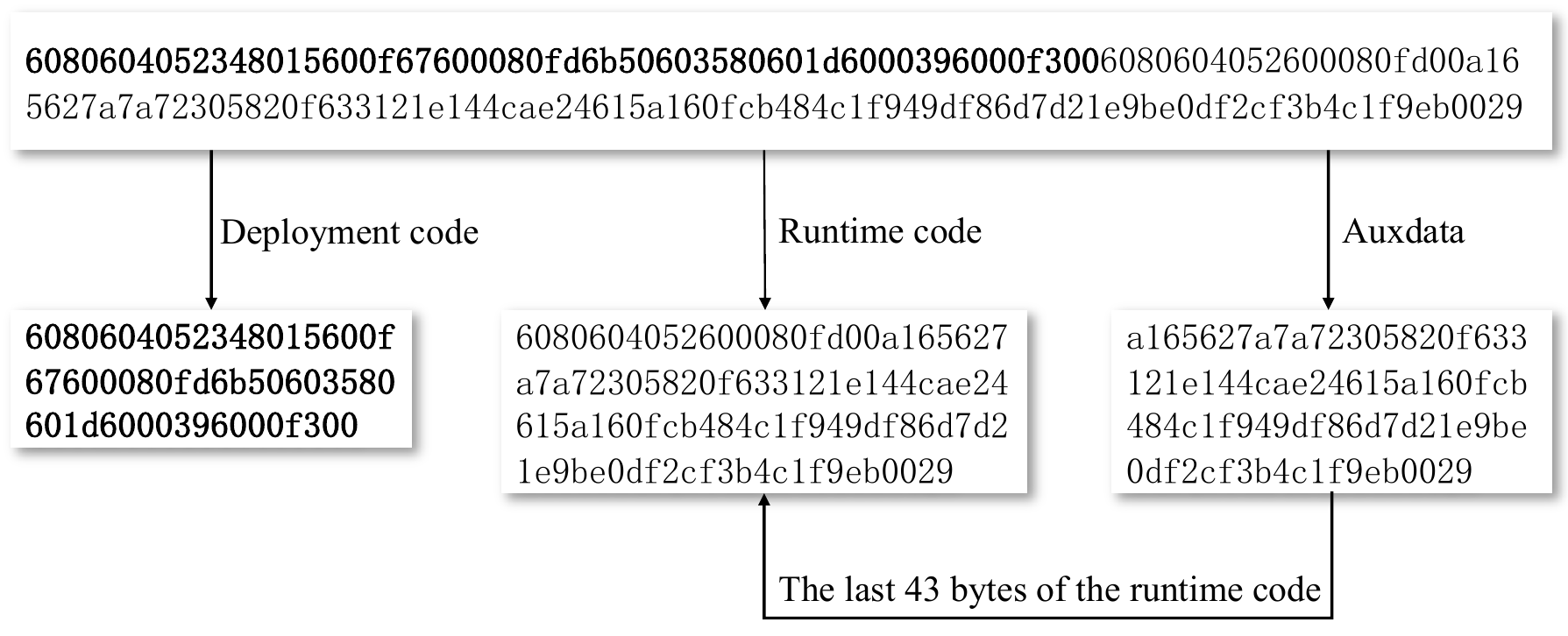}
\caption{The bytecode file of a smart contract consists of three parts: deployment code, runtime code, and auxdata.}
\label{fig:3Bytecode structure}
\end{figure}

The source code of a smart contract is compiled to generate bytecode, which is divided into three parts: deployment code, runtime code, and auxdata. When EVM builds a contract, it first creates the contract user, then runs the deployment code and deposits the two parts, runtime code and auxdata, onto the blockchain, and in the actual operation of the contract, it is the runtime code that runs; the last 43 bytes of each contract are auxdata, which will be saved following the runtime code. An example of bytecode structure is given below as shown in Fig. \ref{fig:3Bytecode structure}.

\subsubsection{Assembly Opcode}
The decompiled code can be obtained by disassembling the bytecode. The decompiled code consists of two parts: the instruction address and the instruction opcode. Since the smart contract only runs the runtime code part when it is executed, the decompilation operation on the bytecode only needs to operate on the runtime code part.

Up to now, EVM has used 145 opcodes, which can be divided into arithmetic operation instructions, comparison operation instructions, per-bit operation instructions, cryptographic calculation instructions, stack, memory, and storage operation instructions, jump instructions, block, and smart contract related instructions, etc. according to their functions. The specific opcodes are divided as shown in Table \ref{Division and function of operation code}.

\begin{table}[htbp]
  \centering
  \caption{The classification of EVM opcodes and the functions of each category, with a few examples.}
  \resizebox{\linewidth}{!}{
    \begin{tabular}{c|c|c}
    \toprule
    \textbf{OPCODE} & \textbf{FUNCTION} & \textbf{EXAMPLE} \\
    \midrule
    0x00 - 0x0B & Stopping and Arithmetic Operation & ADD, SUB, STOP, DIV \\
    0x10 - 0x1A & Comparison and By-bit Logic Operations & GT, LT, EQ \\
    0x20  & Encryption & SHA3 \\
    0x30 - 0x3E & Environmental Information & ADDRESS, CALLER \\
    0x40 - 0x45 & Block Operations & BLOCKHASH, COINBASE \\
    0x50 - 0x5B & Storage and Execution & POP, JUMP, JUMPI \\
    0x50 - 0x5B & Push Operation & PUSH1 - PUSH32 \\
    0x80 - 0x8F & Copy Command & DUP1 - DUP16 \\
    0x90 - 0x9F & Exchange Instructions & SWAP1 - SWAP16 \\
    0xA0 - 0xA4 & Logging Instructions & LOG0 - LOG4 \\
    0xF0 - 0xFF & System Command & CALL, RETURN \\
    \bottomrule
    \end{tabular}%
    }
  \label{Division and function of operation code}%
\end{table}%

An example of decompiling the bytecode is given below, as shown in Fig. \ref{fig:4Decompiling bytecode}.

\begin{figure}[!ht]
\centering
\includegraphics[scale=0.48]{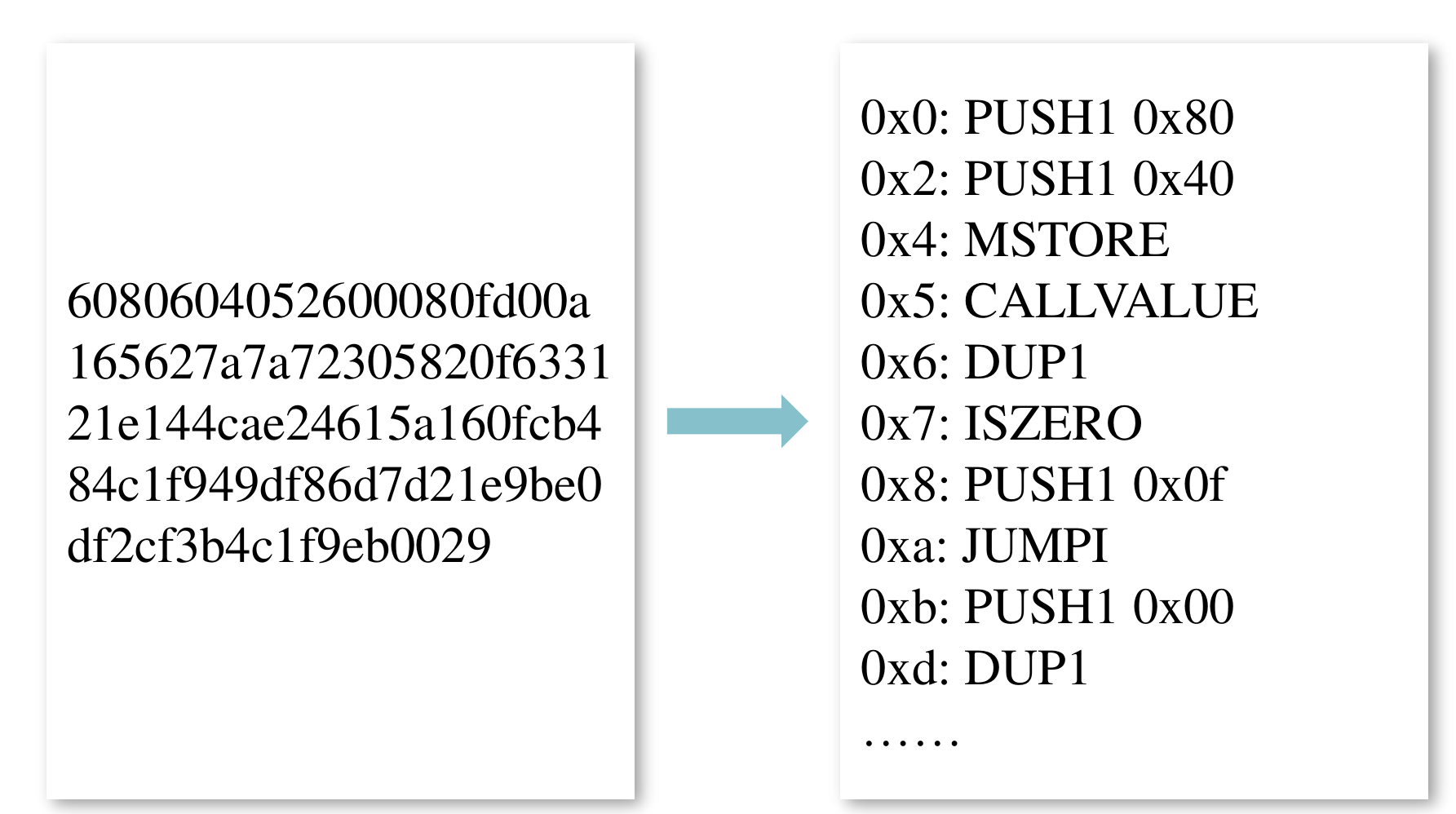}
\caption{An example of decompiling the bytecode file of a smart contract to get the opcode.}
\label{fig:4Decompiling bytecode}
\end{figure}

\subsection{Control Flow Graph Generation}
Building a CFG using the bytecode of a smart contract involves the following main steps.
\begin{enumerate}
    \item Disassembling the hexadecimal bytecode file to obtain the corresponding assembly opcode.
    \item Dividing the opcode into some basic blocks according to the rules for building basic blocks.
    \item Calculating the destination address of each basic block according to transfer instructions such as jump instructions and conditional instructions, and adding edges between the corresponding two basic blocks, thus completing the construction of the control flow graph (CFG).
    \item Based on CFG, sequential dependent edges are added between the sequentially executed basic blocks to improve the graph structure.
\end{enumerate}

The above section has analyzed the bytecode of the smart contract and described how to get the opcode. The next section describes how to build the CFG.

\subsubsection{Basic Block Division}
A basic block is a maximized sequence of instructions in which the execution of an instruction can only start from the first instruction and end with the last instruction. A code file can generate a graph structure by dividing the basic blocks and adding jump dependencies and sequential dependencies.

The following are three basic principles for constructing a basic block.

\begin{enumerate}
    \item If this instruction is the first instruction of a program or subroutine, the current basic block should be terminated and a new basic block should be opened with this instruction as the first instruction in it.
    \item If this instruction is a jump statement or branch statement, etc., the instruction should be used as the last instruction of the current basic block, and then the basic block should be terminated.
    \item If the instruction does not belong to the above two cases, it is added directly to the current basic block.
\end{enumerate}

An example of a bytecode file divided into basic blocks is given below, as shown in Fig. \ref{fig:6Dividing blocks}.

\begin{figure}[!ht]
\centering
\includegraphics[scale=0.32]{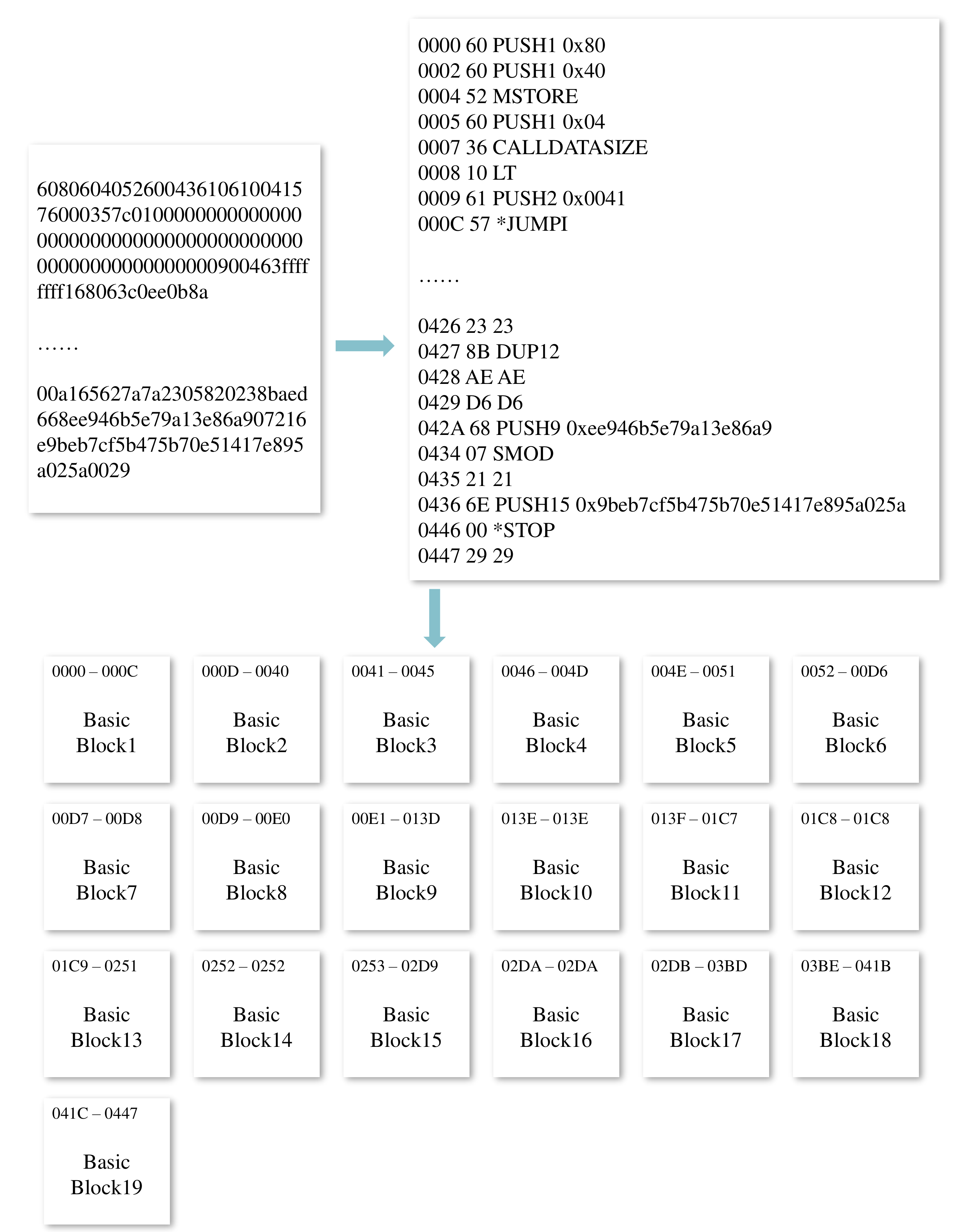}
\caption{According to the three principles of dividing basic blocks, the opcode obtained by decompiling can be divided into several basic blocks.}
\label{fig:6Dividing blocks}
\end{figure}

\subsubsection{CFG Structure Construction}

\begin{figure}[!ht]
\centering
\includegraphics[scale=0.33]{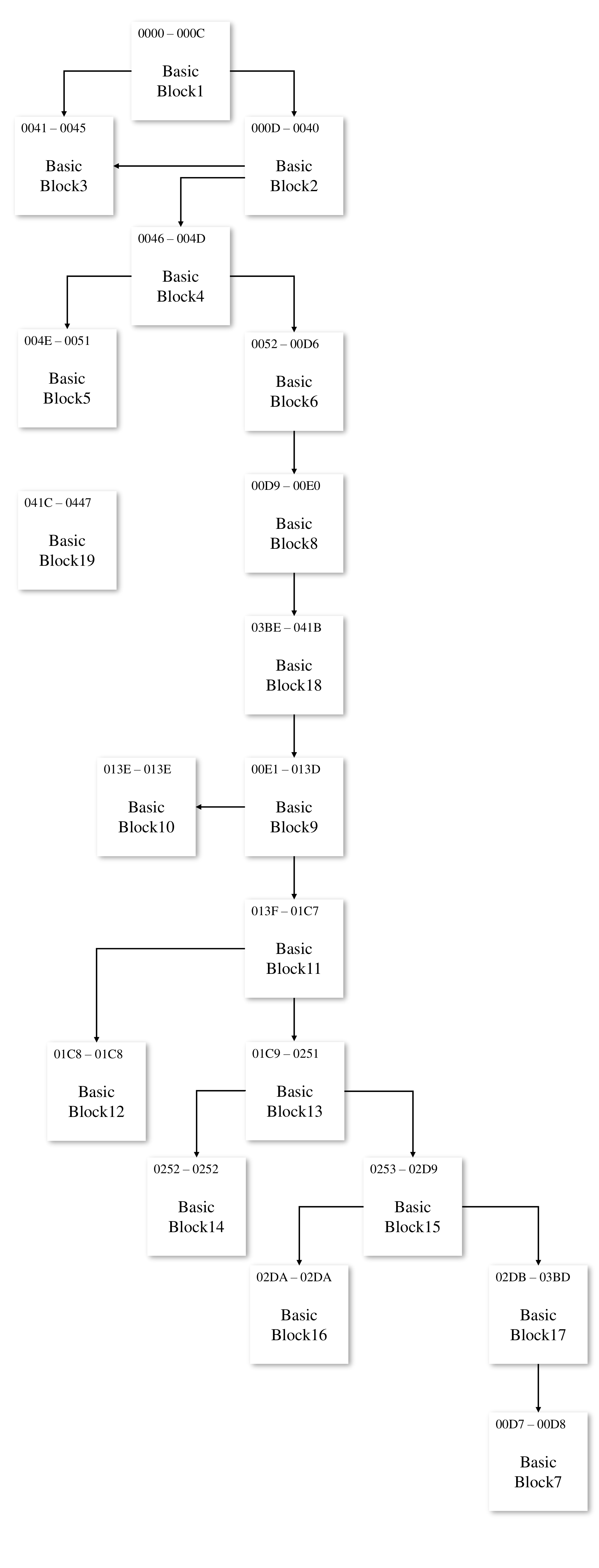}
\caption{After getting the basic blocks, we first add jump edges between the basic blocks according to the jump logic of the opcode, and then add sequential edges between the basic blocks according to the code running flow.}
\label{fig:7CFG}
\end{figure}

After the work of dividing the basic blocks is completed, it is necessary to add new edges to the basic blocks in combination with assembly instructions, i.e., the jumping relationships between the basic blocks. The complete diagram structure after adding sequential edges to the basic blocks divided in the above section is shown in Fig. \ref{fig:7CFG}.

\subsection{GCN Model }

\subsubsection{Convolutional Layer Definition}

The underlying equation of GCN is shown in equation \ref{fig: underlying equation of GCN}.

\begin{equation}
\label{fig: underlying equation of GCN}
H^{l+1} = \sigma(\tilde{D}^{-\frac{1}{2}}\tilde{A}\tilde{D}^{-\frac{1}{2}}H^{l}w^l)
\end{equation}

where $H^l$ is the input feature of the lth layer and $H^{l+1}$ is the output feature. $w^l$ is the linear transformation matrix, i.e., the weight matrix that the model needs to learn, and $\sigma(\cdot)$ is the nonlinear activation function, such as ReLU, Sigmoid, etc.

$\tilde{A}$ is the adjacency matrix with self-connections (hereafter referred to as the self-connected adjacency matrix), defined as shown in equation \ref{defination of A}.

\begin{equation}
\label{defination of A}
    \tilde{A} = A + I
\end{equation}

$A$ is the adjacency matrix and $I$ is the unit matrix. In the adjacency matrix, the elements at the diagonal positions represent the relationship between the node and itself, while the elements at the non-diagonal positions represent the relationship between the node and the node. If a node is not connected to itself, the element at the diagonal position is 0. However, such a setting will cause problems in subsequent calculations, i.e., it is impossible to distinguish between "own nodes" and "unconnected nodes" (both of which have the corresponding element position of 0).

\begin{figure}[!ht]
\centering
\includegraphics[scale=0.72]{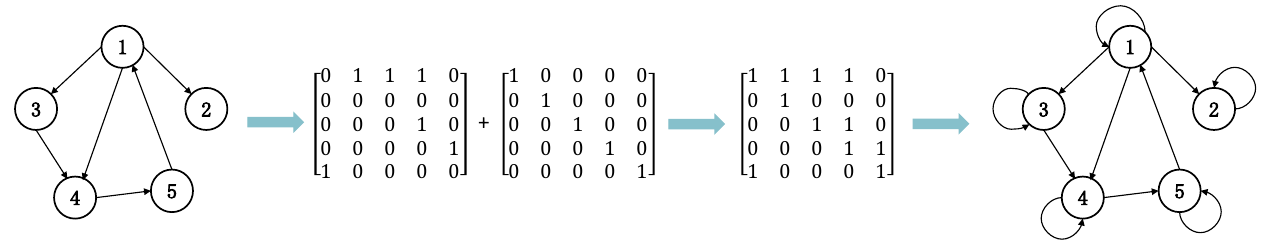}
\caption{Definition of $\tilde{A}$ matrix.}
\label{fig:1Definition of A matrix}
\end{figure}

An example is given below to better illustrate the definition of the  matrix, as shown in Fig. \ref{fig:1Definition of A matrix}.

$\tilde{D}$ is the degree matrix of the self-connected matrix, defined as shown in equation \ref{Defination of D}.

\begin{equation}
\label{Defination of D}
    \tilde{D_{ij}} = \sum_j\tilde{A}_{ij}
\end{equation}

The definition of the degree matrix is still illustrated below using the data in Fig. \ref{fig:1Definition of A matrix}, where $\tilde{D}^{-\frac{1}{2}}$ is the inverse of the square root taken from the basis of the self-connected degree matrix, as shown in equation \ref{D}.

\begin{equation}
\resizebox{0.9\hsize}{!}{$
\label{D}
\tilde{A}  = \begin{bmatrix}
  1&  1&  1&  1& 0\\
  0&  1&  0&  0& 0\\
  0&  0&  1&  1& 0\\
  0&  0&  0&  1& 1\\
  1&  0&  0&  0& 1
\end{bmatrix}
\quad
\tilde{D}  = \begin{bmatrix}
  4&  0&  0&  0& 0\\
  0&  1&  0&  0& 0\\
  0&  0&  2&  0& 0\\
  0&  0&  0&  2& 0\\
  0&  0&  0&  0& 2
\end{bmatrix}
\tilde{D}^{-\frac{1}{2} } = \begin{bmatrix}
  1&  1&  1&  1& 0\\
  0&  1&  0&  0& 0\\
  0&  0&  1&  1& 0\\
  0&  0&  0&  1& 1\\
  1&  0&  0&  0& 1
\end{bmatrix}$}
\end{equation}

\subsubsection{GCN Model Definition}

\begin{figure}[!ht]
\centering
\includegraphics[scale=0.3]{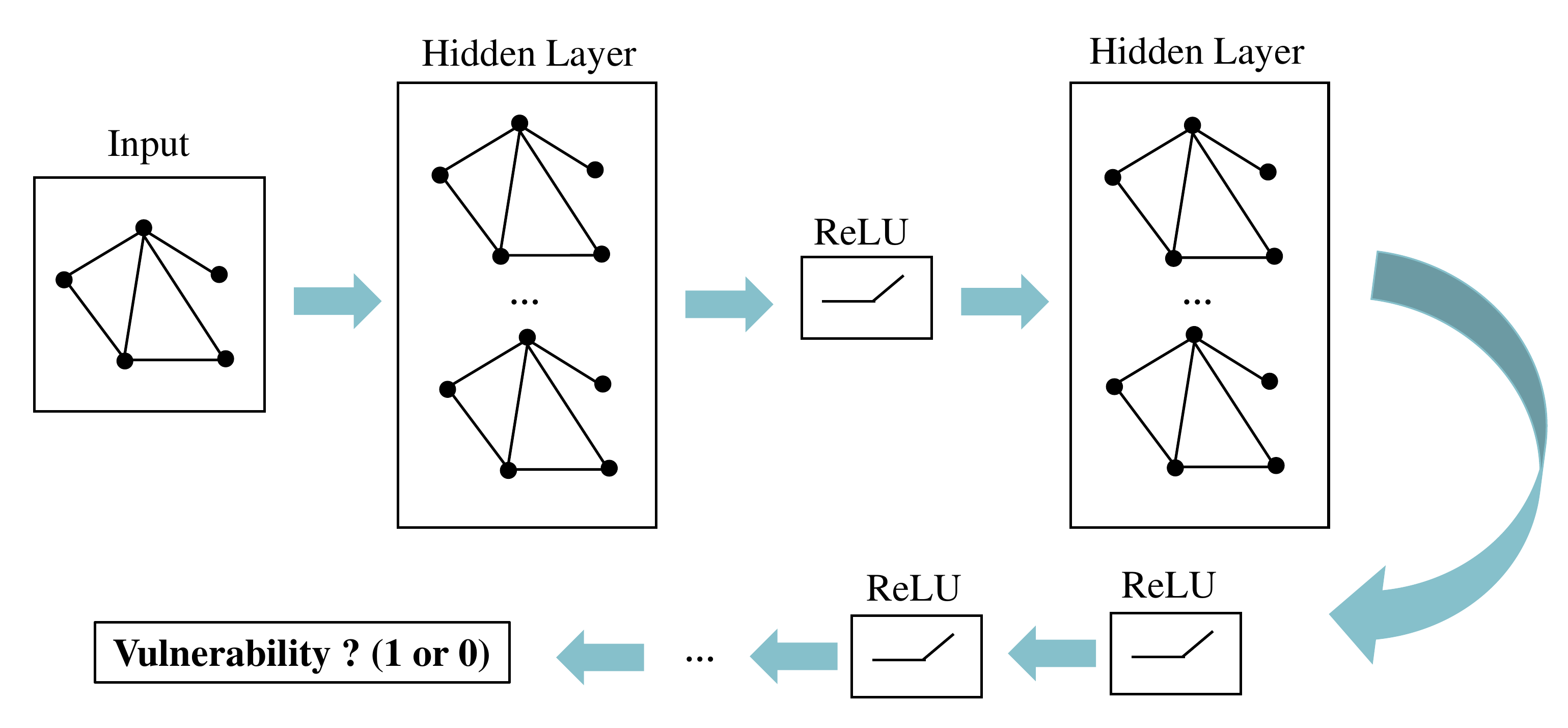}
\caption{The GCN model structure used in this paper uses a network model with different layers of hidden layers to predict contract vulnerabilities.}
\label{fig:9Network model}
\end{figure}

The model is used to predict the label $\hat{y}$, when $\hat{y}=1$, it indicates that there is some vulnerability, otherwise, it indicates that the smart contract is secure. The network model is described below, the specific network model is shown in Fig. \ref{fig:9Network model}.

The network model consists of an input layer, an output layer, and some hidden layers, where each layer is computed and the results are fed into the activation function $ReLU(\cdot)$. After several layers of computation, a prediction label is an output by the output layer, where 1 indicates that the contract has some kind of vulnerability, otherwise it indicates that the contract is secure.

The process of CFG construction has been described in the previous section, using the adjacency matrix A and the node feature matrix X to represent the corresponding CFG as the input to the network model. Since the work in this paper does not involve natural semantic processing for the operand part, the node feature matrix X will be used instead of the unit matrix.

In this paper, we try to detect smart contracts using a network model containing hidden layers from 1 to 4 layers, examine the effect of the number of network layers on each evaluation metric, and analyze the reasons for metric changes.

\section{Experimental Results and Analysis}

In this section, we first introduce the dataset for the experiments, and then describe the details and results of the experiments.

\subsection{Smart Contracts Dataset}
The current public dataset of smart contracts is in the form of source code, so you need to compile the smart contract source code file based on the public dataset and get the smart contract bytecode file according to the compiler version declared inside the contract. It should also be noted that different versions of smart contract compilers are not compatible with each other, so the compilation process should strictly follow the declared compiler version to avoid problems due to the compiler version.

In this paper, we use the publicly available source code dataset for compilation, produce a dataset containing 1420 bytecodes, and assign a label to each data (set to 1 for the existence of vulnerabilities, otherwise set to 0), among which 472 contain timestamp-dependent vulnerabilities. The dataset was divided into a training set and a test set according to 8:2.

\subsection{Experimental Results}
In this paper, four metrics are used to judge the effectiveness of the model for vulnerability prediction, namely Accuracy, Recall, Precision, and F1-score. TP, FN, FP, and TN are used to represent the classification of the prediction results, where TP denotes contracts that detect the presence of vulnerabilities but have vulnerabilities, FN denotes contracts that detect no exist but have vulnerabilities, FP denotes contracts that are detected to have vulnerabilities but do not have vulnerabilities, and TN denotes contracts that are detected not to have vulnerabilities but do not have vulnerabilities.

Accuracy represents the ratio of the number of correctly detected contracts to the number of all contracts and is calculated as shown in equation \ref{accuracy}.

\begin{equation}
\label{accuracy}
    Accuracy = \frac{TP+TN}{TP+FN+FP+TN}
\end{equation}

Recall represents the ratio of the number of contracts detected with vulnerabilities to the number of all contracts containing vulnerabilities, and is calculated as shown in equation \ref{recall}.

\begin{equation}
\label{recall}
    Recall = \frac{TP}{TP+FN}
\end{equation}

Precision represents the ratio of the number of contracts detected as vulnerable and having vulnerabilities to the number of all contracts detected as containing vulnerabilities and is calculated as shown in equation \ref{precision}.

\begin{equation}
\label{precision}
    Precision = \frac{TP}{TP+FP}
\end{equation}

The F1-score is a comprehensive assessment metric that balances accuracy and recall and can be considered as the inverse average of accuracy and recall, which is calculated as shown in equation \ref{f1-score}.

\begin{equation}
\label{f1-score}
    F1-score = 2 * \frac{Presicion*Recall}{Presicion+Recall}
\end{equation}

\begin{figure}[!ht]
\centering
\includegraphics[scale=0.51]{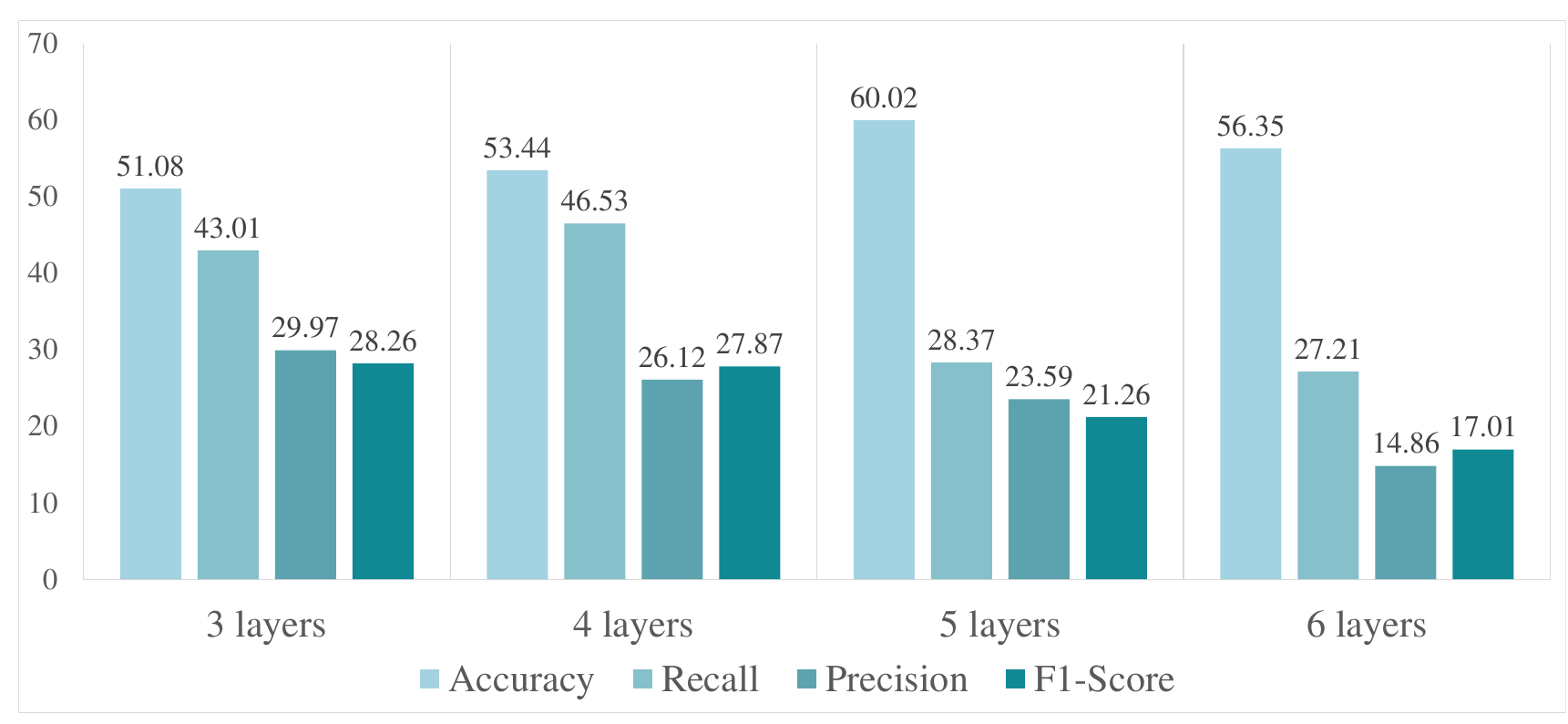}
\caption{Results of ablation experiments.}
\label{fig:Result}
\end{figure}

During the experiment, the number of layers of the GCN model was changed to observe the changes of each index, and the results are shown in Fig. \ref{fig:Result} (for the convenience of graphing, the "result*100" is done on each result).

From the experimental results, it is clear that the accuracy and F1 scores show an overall decreasing trend as the number of network layers increases. And it is easy to observe that there is a greater decrease in recall in the network models with 5 and 6 layers; there is also a small decrease in accuracy in the network model with 6 layers.

According to the structure of neural networks, it is known that the more layers of hidden layers, in addition to the input and output layers, the more significant the non-linearity is. In the process of model learning, which is the process of adjusting and optimizing the weights and thresholds of each connection, the neurons in the latter layer receive the abstract data from the processed neurons in the previous layer, so the higher the number of layers of the network model, the higher it's level of abstraction, and it will show better results on some specific tasks. However, in this problem, it is obvious that an excessively deep network level is not needed, and it is clear from the experimental results that the prediction results of a 3 or 4-layer network are more informative.

In addition, the work in this paper does not incorporate the semantic processing part, and the features of the graph nodes are not well characterized, which is guessed to be the reason for the low precision and F1-Score. The following conjectures may improve the accuracy of the model's prediction at present.

\begin{enumerate}
    \item Add the semantic feature processing part. After decompiling to get the decompiled code, adding the part of natural language processing can get more optimized node feature data and make the prediction results more accurate.
    \item Further classify the edges of the control flow graph, for example, they can be divided into conditional jumping edges, and sequential jumping edges, to optimize the feature data.
\end{enumerate}

The current existing research has uneven work in the part of generating feature data, and the node feature data can describe the node meaning, which has a relatively large impact on the model prediction results.

\subsection{Experiment Comparison}

This paper lists the results of other vulnerability detection methods for smart contracts, of which there are three based on non-deep learning, respectively, a smart contract automatic audit tool oyente \cite{10.1145/2976749.2978309}, an inspection tool based on symbolic execution techniques Mythril \cite{web1}, and a static analysis tool Smart check \cite{8445052}; two deep learning based methods LSTM and GRU. The specific comparison results are shown in Table \ref{comparison}.

\begin{table}[!ht]
\caption{The results of the listed non-deep learning based methods and deep learning based methods for contract vulnerability detection in Accuracy, Recall, Presicion and F1-Score for comparison with the method GCN in this paper.}
\label{comparison}
\resizebox{\linewidth}{!}{
\begin{tabular}{cc|cccc}
\hline
\multicolumn{2}{c|}{\textbf{Methods}}                                           & \textbf{Accuracy} & \textbf{Recall} & \textbf{Precision} & \textbf{F1-Score} \\ \hline
\multicolumn{1}{c|}{\multirow{3}{*}{\textbf{Non-deep Learning}}} & Oyente       & 59.45             & 38.44           & 45.16              & 41.53             \\
\multicolumn{1}{c|}{}                                            & Mythril      & 61.08             & 41.72           & 50.00                 & 45.49             \\
\multicolumn{1}{c|}{}                                            & Smart check  & 44.32             & 37.25           & 39.16              & 38.18             \\ \hline
\multicolumn{1}{c|}{\multirow{3}{*}{\textbf{Deep learning}}}     & LSTM         & 50.79             & 59.23           & 50.23              & 54.41             \\
\multicolumn{1}{c|}{}                                            & GRU          & 52.06             & 59.91           & 49.41              & 54.15             \\
\multicolumn{1}{c|}{}                                            & \textbf{Our method} & \textbf{53.44}    & \textbf{46.53}  & \textbf{26.12}     & \textbf{27.87}    \\ \hline
\end{tabular}}
\end{table}

Among these methods, the one that can achieve the best results is Mythril, with an accuracy rate of 61.08\%. The reason is that its detection principle is relatively complex and requires taint analysis and other related technologies, so it can achieve better results. The accuracy of deep learning-based vulnerability detection methods is slightly above 50\%, but the GCN model used in this paper has better results. It has promising results under the conditions of using the basic network model and the basic composition strategy, which is sufficient to show the effectiveness of using smart contract bytecode files for vulnerability detection and the GCN model. Optimizing feature data and network models on this basis is more likely to result in better data.

\section{Conclusion}
This paper introduces a method of applying graph neural networks to smart contract vulnerability detection, and the experimental results show that vulnerability detection using bytecode is a feasible detection method. When constructing the network model, it is important to choose the appropriate network depth and not blindly increase the number of hidden layers. Subsequent research of such work should focus on how to generate graph structures, whether using smart contract source code or smart contract bytecode, the feature data should be able to better express the invocation relationship between functions, the execution process of the contract, and the semantics of the contract instructions. On this basis with a suitable graph neural network model, the prediction results can be further optimized.

\section*{Acknowledgments}
The research work of this paper were supported by the National Natural Science Foundation of China (No. 62177022, 61901165, 61501199), Collaborative Innovation Center for Informatization and Balanced Development of K-12 Education by MOE and Hubei Province (No. xtzd2021-005), and Self-determined Research Funds of CCNU from the Colleges’ Basic Research and Operation of MOE (No. CCNU22QN013).

\bibliography{myref,Citations}{}     

\begin{thebibliography}{10}
\providecommand{\url}[1]{#1}
\csname url@samestyle\endcsname
\providecommand{\newblock}{\relax}
\providecommand{\bibinfo}[2]{#2}
\providecommand{\BIBentrySTDinterwordspacing}{\spaceskip=0pt\relax}
\providecommand{\BIBentryALTinterwordstretchfactor}{4}
\providecommand{\BIBentryALTinterwordspacing}{\spaceskip=\fontdimen2\font plus
\BIBentryALTinterwordstretchfactor\fontdimen3\font minus
  \fontdimen4\font\relax}
\providecommand{\BIBforeignlanguage}[2]{{%
\expandafter\ifx\csname l@#1\endcsname\relax
\typeout{** WARNING: IEEEtran.bst: No hyphenation pattern has been}%
\typeout{** loaded for the language `#1'. Using the pattern for}%
\typeout{** the default language instead.}%
\else
\language=\csname l@#1\endcsname
\fi
#2}}
\providecommand{\BIBdecl}{\relax}
\BIBdecl

\bibitem{9468408}
M.~A. Rahman, M.~S. Abuludin, L.~X. Yuan, M.~S. Islam, and A.~T. Asyhari,
  ``Educhain: Cia-compliant blockchain for intelligent cyber defense of
  microservices in education industry 4.0,'' \emph{IEEE Transactions on
  Industrial Informatics}, vol.~18, no.~3, pp. 1930--1938, 2022.

\bibitem{Lyu2022}
L.~Lyu, Z.~Wang, H.~Yun, Z.~Yang, and Y.~Li, ``Deep knowledge tracing based on
  spatial and temporal representation learning for learning performance
  prediction,'' \emph{Applied Sciences}, vol.~12, no.~14, pp. 1--21, Jan. 2022.

\bibitem{9459573}
Z.~Li and Z.~Ma, ``A blockchain-based credible and secure education experience
  data management scheme supporting for searchable encryption,'' \emph{China
  Communications}, vol.~18, no.~6, pp. 172--183, 2021.

\bibitem{8247166}
M.~Turkanović, M.~Hölbl, K.~Košič, M.~Heričko, and A.~Kamišalić,
  ``Eductx: A blockchain-based higher education credit platform,'' \emph{IEEE
  Access}, vol.~6, pp. 5112--5127, 2018.

\bibitem{visualanalysis}
Y.~Zhao and X.~Zhao, ``\BIBforeignlanguage{Chinese;}{A visual analysis of
  blockchain applications in education},''
  \emph{\BIBforeignlanguage{Chinese;}{Journal of the Hebei Academy of
  Sciences}}, vol.~39, no.~03, pp. 14--20, 2022.

\bibitem{Zeng2022}
C.~Zeng, J.~Ye, Z.~Wang, N.~Zhao, and M.~Wu, ``Cascade neural network-based
  joint sampling and reconstruction for image compressed sensing,''
  \emph{Signal, Image and Video Processing}, vol.~16, no.~1, pp. 47--54, Feb.
  2022.

\bibitem{9893128}
P.~Ocheja, F.~J. Agbo, S.~S. Oyelere, B.~Flanagan, and H.~Ogata, ``Blockchain
  in education: A systematic review and practical case studies,'' \emph{IEEE
  Access}, vol.~10, pp. 99\,525--99\,540, 2022.

\bibitem{Wang2022ac}
Z.~Wang, Z.~Wang, C.~Zeng, Y.~Yu, and X.~Wan, ``High-quality image compressed
  sensing and reconstruction with multi-scale dilated convolutional neural
  network,'' \emph{Circuits, Systems, and Signal Processing}, Sep. 2022.

\bibitem{__2017}
X.~Yang, X.~Li, H.~Wu, and K.~Zhao, ``\BIBforeignlanguage{Chinese;}{Blockchain
  technology in the field of education application model and real
  challenges},'' \emph{\BIBforeignlanguage{Chinese;}{Modern Distance Education
  Research}}, no.~02, pp. 34--45, 2017.

\bibitem{Wang2022t}
Z.~Wang, Y.~Yang, C.~Zeng, S.~Kong, S.~Feng, and N.~Zhao, ``Shallow and deep
  feature fusion for digital audio tampering detection,'' \emph{EURASIP Journal
  on Advances in Signal Processing}, vol. 2022, no.~1, p.~69, Aug. 2022.

\bibitem{8246573}
T.~T.~A. Dinh, R.~Liu, M.~Zhang, G.~Chen, B.~C. Ooi, and J.~Wang, ``Untangling
  blockchain: A data processing view of blockchain systems,'' \emph{IEEE
  Transactions on Knowledge and Data Engineering}, vol.~30, no.~7, pp.
  1366--1385, 2018.

\bibitem{Zeng2022a}
C.~Zeng, Y.~Yang, Z.~Wang, S.~Kong, and S.~Feng, ``Audio tampering forensics
  based on representation learning of enf phase sequence,'' \emph{International
  Journal of Digital Crime and Forensics}, vol.~14, no.~1, pp. 1--19, Jan.
  2022.

\bibitem{10.1145/2976749.2978309}
\BIBentryALTinterwordspacing
L.~Luu, D.-H. Chu, H.~Olickel, P.~Saxena, and A.~Hobor, ``Making smart
  contracts smarter,'' in \emph{Proceedings of the 2016 ACM SIGSAC Conference
  on Computer and Communications Security}, ser. CCS '16.\hskip 1em plus 0.5em
  minus 0.4em\relax New York, NY, USA: Association for Computing Machinery,
  2016, p. 254–269. [Online]. Available:
  \url{https://doi.org/10.1145/2976749.2978309}
\BIBentrySTDinterwordspacing

\bibitem{10.1145/3238147.3238177}
\BIBentryALTinterwordspacing
B.~Jiang, Y.~Liu, and W.~K. Chan, ``Contractfuzzer: Fuzzing smart contracts for
  vulnerability detection,'' in \emph{Proceedings of the 33rd ACM/IEEE
  International Conference on Automated Software Engineering}, ser. ASE
  2018.\hskip 1em plus 0.5em minus 0.4em\relax New York, NY, USA: Association
  for Computing Machinery, 2018, p. 259–269. [Online]. Available:
  \url{https://doi.org/10.1145/3238147.3238177}
\BIBentrySTDinterwordspacing

\bibitem{Zeng2021a}
C.~Zeng, D.~Zhu, Z.~Wang, M.~Wu, W.~Xiong, and N.~Zhao, ``Spatial and temporal
  learning representation for end-to-end recording device identification,''
  \emph{EURASIP Journal on Advances in Signal Processing}, vol. 2021, no.~1,
  p.~41, 2021.

\bibitem{__2021-1273}
C.~Shen, ``\BIBforeignlanguage{Chinese;}{Research on smart contract
  vulnerability detection method based on deep learning},'' Master, Wuhan
  University, 2021.

\bibitem{Zeng2020}
C.~Zeng, D.~Zhu, Z.~Wang, Z.~Wang, N.~Zhao, and L.~He, ``An end-to-end deep
  source recording device identification system for web media forensics,''
  \emph{International Journal of Web Information Systems}, vol.~16, no.~4, pp.
  413--425, Aug. 2020.

\bibitem{tann_towards_2019}
\BIBentryALTinterwordspacing
W.~J.-W. Tann, X.~J. Han, S.~S. Gupta, and Y.-S. Ong,
  ``\BIBforeignlanguage{en}{Towards {Safer} {Smart} {Contracts}: {A} {Sequence}
  {Learning} {Approach} to {Detecting} {Security} {Threats}},'' Jun. 2019,
  arXiv:1811.06632 [cs]. [Online]. Available:
  \url{http://arxiv.org/abs/1811.06632}
\BIBentrySTDinterwordspacing

\bibitem{TextClassification}
Y.~Tan, J.~Wang, and C.~Zhang, ``\BIBforeignlanguage{Chinese;}{A review of text
  classification methods based on graph convolutional neural networks},''
  \emph{\BIBforeignlanguage{Chinese;}{Computer Science}}, vol.~49, no.~08, pp.
  205--216, 2022.

\bibitem{Zeng2022b}
C.~Zeng, K.~Yan, Z.~Wang, Y.~Yu, S.~Xia, and N.~Zhao, ``Abs-cam: A gradient
  optimization interpretable approach for explanation of convolutional neural
  networks,'' \emph{Signal, Image and Video Processing}, pp. 1--8, Jul. 2022.

\bibitem{Wang2021}
Z.~Wang, C.~Zuo, and C.~Zeng, ``Sae based unified double jpeg compression
  detection system for web image forensics,'' \emph{International Journal of
  Web Information Systems}, vol.~17, no.~2, pp. 84--98, Apr. 2021.

\bibitem{Wang2023}
Z.~Wang, W.~Yan, C.~Zeng, Y.~Tian, and S.~Dong, ``A unified interpretable
  intelligent learning diagnosis framework for learning performance prediction
  in intelligent tutoring systems,'' \emph{International Journal of Intelligent
  Systems}, vol. 2023, pp. 1--20, 2023.

\bibitem{Li2023a}
L.~Li and Z.~Wang, ``Calibrated q-matrix-enhanced deep knowledge tracing with
  relational attention mechanism,'' \emph{Applied Sciences}, vol.~13, no.~4,
  pp. 1--24, Jan. 2023.

\bibitem{Min2019}
Q.~Min, Z.~Wang, and N.~Liu, ``Integrating a cloud learning environment into
  english-medium instruction to enhance non-native english-speaking students'
  learning,'' \emph{Innovations in Education and Teaching International},
  vol.~56, no.~4, pp. 493--504, Jul. 2019.

\bibitem{Wang2022as}
Z.~Wang, W.~Wu, C.~Zeng, J.~Yao, Y.~Yang, and H.~Xu, ``Smart contract
  vulnerability detection for educational blockchain based on graph neural
  networks,'' in \emph{2022 International Conference on Intelligent Education
  and Intelligent Research (IEIR)}, Dec. 2022, pp. 8--14.

\bibitem{Zeng2021c}
C.~Zeng, Z.~Wang, Z.~Wang, K.~Yan, and Y.~Yu, ``Image compressed sensing and
  reconstruction of multi-scale residual network combined with channel
  attention mechanism,'' \emph{Journal of Physics: Conference Series}, vol.
  2010, no.~1, p. 012134, Sep. 2021.

\bibitem{Li2023}
L.~Li, Z.~Wang, and T.~Zhang, ``Gbh-yolov5: Ghost convolution with
  bottleneckcsp and tiny target prediction head incorporating yolov5 for pv
  panel defect detection,'' \emph{Electronics}, vol.~12, no.~3, pp. 1--15, Jan.
  2023.

\bibitem{Wang2017}
Z.-F. Wang, L.~Zhu, Q.-S. Min, and C.-Y. Zeng, ``Double compression detection
  based on feature fusion,'' in \emph{2017 International Conference on Machine
  Learning and Cybernetics (ICMLC)}.\hskip 1em plus 0.5em minus 0.4em\relax
  Ningbo: IEEE, Jul. 2017, pp. 379--384.

\bibitem{Wang2015a}
Z.~Wang, Q.~Liu, H.~Yao, and J.~Chen, ``Virtual chime-bells experimental system
  based on multi-modal fusion,'' in \emph{2015 International Conference of
  Educational Innovation through Technology (EITT)}.\hskip 1em plus 0.5em minus
  0.4em\relax Wuhan, China: IEEE, Oct. 2015, pp. 64--67.

\bibitem{Zeng2020a}
C.~Zeng, Z.~Wang, and Z.~Wang, ``Image reconstruction of iot based on parallel
  cnn,'' in \emph{2020 International Conferences on Internet of Things
  (iThings)}.\hskip 1em plus 0.5em minus 0.4em\relax Rhodes, Greece: IEEE, Nov.
  2020, pp. 258--263.

\bibitem{Tian2018a}
Y.~Tian, X.~Wang, H.~Yao, J.~Chen, Z.~Wang, and L.~Yi, ``Occlusion handling
  using moving volume and ray casting techniques for augmented reality
  systems,'' \emph{Multimedia Tools and Applications}, vol.~77, no.~13, pp.
  16\,561--16\,578, Jul. 2018.

\bibitem{Min2018}
Q.~Min, Z.~Wang, and N.~Liu, ``An evaluation of html5 and webgl for medical
  imaging applications,'' \emph{Journal of Healthcare Engineering}, vol. 2018,
  p. e1592821, Aug. 2018.

\bibitem{Wang2022at}
Z.~Wang, J.~Yao, C.~Zeng, W.~Wu, H.~Xu, and Y.~Yang, ``Yolov5 enhanced learning
  behavior recognition and analysis in smart classroom with multiple
  students,'' in \emph{2022 International Conference on Intelligent Education
  and Intelligent Research (IEIR)}, Dec. 2022, pp. 23--29.

\bibitem{Wang2021m}
Z.~Wang, C.~Zeng, S.~Duan, H.~Ouyang, and H.~Xu, ``Robust speaker recognition
  based on stacked auto-encoders,'' in \emph{Advances in Networked-Based
  Information Systems}, L.~Barolli, K.~F. Li, T.~Enokido, and M.~Takizawa,
  Eds.\hskip 1em plus 0.5em minus 0.4em\relax Cham: Springer International
  Publishing, 2021, vol. 1264, pp. 390--399.

\bibitem{Wang2020h}
Z.~Wang, S.~Duan, C.~Zeng, X.~Yu, Y.~Yang, and H.~Wu, ``Robust speaker
  identification of iot based on stacked sparse denoising auto-encoders,'' in
  \emph{2020 International Conferences on Internet of Things (iThings)}.\hskip
  1em plus 0.5em minus 0.4em\relax Rhodes, Greece: IEEE, Nov. 2020, pp.
  252--257.

\bibitem{Zeng2021b}
C.~Zeng, D.~Zhu, Z.~Wang, and Y.~Yang, ``Deep and shallow feature fusion and
  recognition of recording devices based on attention mechanism,'' in
  \emph{Advances in Intelligent Networking and Collaborative Systems},
  L.~Barolli, K.~F. Li, and H.~Miwa, Eds.\hskip 1em plus 0.5em minus
  0.4em\relax Cham: Springer International Publishing, 2021, vol. 1263, pp.
  372--381.

\bibitem{Wang2018a}
Z.-F. Wang, J.~Wang, C.-Y. Zeng, Q.-S. Min, Y.~Tian, and M.-Z. Zuo, ``Digital
  audio tampering detection based on enf consistency,'' in \emph{2018
  International Conference on Wavelet Analysis and Pattern Recognition
  (ICWAPR)}.\hskip 1em plus 0.5em minus 0.4em\relax Chengdu: IEEE, Jul. 2018,
  pp. 209--214.

\bibitem{Zeng2018}
C.-Y. Zeng, C.-F. Ma, Z.-F. Wang, and J.-X. Ye, ``Stacked autoencoder networks
  based speaker recognition,'' in \emph{2018 International Conference on
  Machine Learning and Cybernetics (ICMLC)}.\hskip 1em plus 0.5em minus
  0.4em\relax Chengdu: IEEE, Jul. 2018, pp. 294--299.

\bibitem{Wang2015b}
Z.~Wang, Q.~Liu, J.~Chen, and H.~Yao, ``Recording source identification using
  device universal background model,'' in \emph{2015 International Conference
  of Educational Innovation through Technology (EITT)}.\hskip 1em plus 0.5em
  minus 0.4em\relax Wuhan, China: IEEE, Oct. 2015, pp. 19--23.

\bibitem{Zhu2013}
Z.-Y. Zhu, Q.-H. He, X.-H. Feng, Y.-X. Li, and Z.-F. Wang, ``Liveness detection
  using time drift between lip movement and voice,'' in \emph{2013
  International Conference on Machine Learning and Cybernetics}, vol.~02, Jul.
  2013, pp. 973--978.

\bibitem{Wang2011}
Z.-F. Wang, G.~Wei, and Q.-H. He, ``Channel pattern noise based playback attack
  detection algorithm for speaker recognition,'' in \emph{2011 International
  Conference on Machine Learning and Cybernetics}, vol.~4, Jul. 2011, pp.
  1708--1713.

\bibitem{Wang2011a}
Z.-F. Wang, Q.-H. He, X.-Y. Zhang, H.-Y. Luo, and Z.-S. Su, ``Playback attack
  detection based on channel pattern noise,'' \emph{Journal of South China
  University of Technology}, vol.~39, no.~10, pp. 7--12, 2011.

\bibitem{web1}
\BIBentryALTinterwordspacing
B.~Mueller, ``A framework for bug hunting on the ethereum blockchain,'' 2017.
  [Online]. Available: \url{https: //github.com/ConsenSys/mythril}
\BIBentrySTDinterwordspacing

\bibitem{8445052}
S.~Tikhomirov, E.~Voskresenskaya, I.~Ivanitskiy, R.~Takhaviev, E.~Marchenko,
  and Y.~Alexandrov, ``Smartcheck: Static analysis of ethereum smart
  contracts,'' in \emph{2018 IEEE/ACM 1st International Workshop on Emerging
  Trends in Software Engineering for Blockchain (WETSEB)}, 2018, pp. 9--16.

\end{thebibliography}
\bibliographystyle{IEEEtran}

\end{document}